\documentclass[a4paper,nobibnotes,nofootinbib]{revtex4}

\def\ba{\begin{eqnarray}}
\def\ea{\end{eqnarray}}
\def\eaa{\end{array}}
\def\beqar{\begin{array}}
\def\bars{\begin{eqnarray*}}
\def\ears{\end{eqnarray*}}

\def\be{\begin{equation}}
\def\ee{\end{equation}}
\def\f{\frac}
\def\g{\gamma}

\def\ac{\f {\alpha_s N_c}{\pi}}

\def\ts{\tilde{s}}
\def\tc{\tilde{c}}
\def\n{\nonumber \\}

\newcommand{\bt}{\begin{tabular}}
\newcommand{\et}{\end{tabular}}
\newcommand{\bd}{\begin{displaymath}}
\newcommand{\ed}{\end{displaymath}\noindent}
\newcommand{\ec}{\end{center}}
\newcommand{\bc}{\begin{center}}
\newcommand{\ei}{\end{itemize}}
\newcommand{\bi}{\begin{itemize}}
\newcommand{\ii}{\item}

\newenvironment{rmenum}{
  \newcounter{icount}
  \begin{list}{\roman{icount})}{
    \settowidth{\labelwidth}{viii)}
    \setlength{\itemsep}{-2pt}
    \setlength{\topsep}{2pt}
    \setlength{\partopsep}{0pt}
    \setlength{\parskip}{0pt}
    \setlength{\leftmargin}{\labelwidth}
    \usecounter{icount}}
}{
  \end{list}
}

\usepackage{amsmath,amssymb}
\usepackage{epsfig}

\begin{document}
\title{QCD traveling waves at non-asymptotic energies}

\author{C. Marquet}
\email{marquet@spht.saclay.cea.fr}
\author{R. Peschanski}
\email{pesch@spht.saclay.cea.fr}
\author{G. Soyez\footnote{on leave from the fundamental theoretical physics 
group of the University of Li\`ege.}}
\email{gsoyez@spht.saclay.cea.fr}
\affiliation{Service de physique th{\'e}orique, CEA/Saclay,
  91191 Gif-sur-Yvette cedex, France\footnote{%
URA 2306, unit\'e de recherche associ\'ee au CNRS.}}

\begin{abstract}
Using consistent truncations of the BFKL kernel, we derive 
analytical traveling-wave solutions of the Balitsky-Kovchegov saturation 
equation for both fixed and running coupling. A universal parametrization of the 
``interior'' of the wave front is obtained and compares well with numerical 
simulations of the original Balitsky-Kovchegov equation, even at non-asymptotic 
energies. Using this universal parametrization, we find evidence for a 
traveling-wave pattern of the dipole amplitude determined from the gluon 
distribution extracted from deep inelastic scattering data.
\end{abstract}

\maketitle

\section{Introduction}

In the past decade, there has been a large amount of work devoted to the 
description and understanding of quantum chromodynamics (QCD) in the 
high-energy /density limit corresponding to saturation. On the theoretical side, 
progress is made in obtaining non-linear QCD equations describing the evolution 
of scattering amplitudes towards saturation. On the phenomenological side, the 
discovery of geometric scaling \cite{Stasto:2000er} in deep inelastic scattering 
(DIS) at HERA was a stimulating observation. Indeed geometric scaling has a 
natural explanation \cite{Munier:2003vc} in terms of traveling-wave solutions of 
the Balitsky-Kovchegov (BK) non-linear equation \cite{Balitsky,kov}.

The BK equation is an equation for the evolution in rapidity $Y$ of the 
scattering amplitude $N(\mathbf{r},\mathbf{b},Y)$ of a dipole (a colorless 
$q\bar q$ pair) off a given target. $\mathbf{r}$ is the transverse size of the 
dipole and $\mathbf{b}$ its impact parameter. The rapidity $Y$ is the logarithm 
of the squared total center-of-mass energy of the collision. Assuming that the 
amplitude $N$ is $\mathbf{b}-$independent, the BK equation has been shown 
\cite{Munier:2003vc} to lie in the same university class as the 
Fisher-Kolmogorov-Petrovsky-Piscounov (FKPP) equation \cite{KPP}, extensively 
studied in statistical physics. 

The FKPP equation is an evolution equation for the function $u(x,t)$ where $x$ 
is a space variable and $t$ is the time. It admits \cite{Bramson} asymptotic 
solutions which are functions of the sole variable $x\!-\!vt,$ namely 
traveling-wave solutions $u(x\!-\!vt)$ where $v$ is the speed of the wave. It 
comes from a critical regime obtained by the competition of the exponential 
growth in the linear regime and the non-linear damping. This mechanism for the 
formation of traveling waves is more general \cite{ebert} and applies to the BK 
equation for which time is replaced by rapidity and the logarithm of the dipole 
size plays the role of the position. It has even been extended beyond the 
$\mathbf{b}-$independent case \cite{us} to processes with non-zero momentum 
transfer.

However there are limitations for applying these solutions precisely because 
they are asymptotic, i.e. they require very large values of $Y$ to 
appear; also analytical expressions are only known for relatively small values 
of $r\!=\!|\mathbf{r}|$. This leads to a limitation of analytical predictions to 
the tail of the wave ($N\!\ll\!1$) which is essentially determined by the linear 
regime. For instance, the ``interior'' of the wave is not reproduced; it 
corresponds to the transition to saturation, a intermediate regime between the 
tail ($N\!\ll\!1$) and the saturated region ($N\!\sim\!1$). That moderates 
the phenomenological impact of those mathematically powerful results. 

In order to circumvent these difficulties, it is necessary to extend the study 
to the non-asymptotic regime in a region of larger $N,$ which implies to use 
more information on the non-linear term of the equation than the previous 
approaches. For this sake, a new approach to study the BK equation has been 
developed in \cite{robi}. In short, the guiding idea is to try and find exact 
solutions of approximate QCD equations rather than approximate solutions of 
exact QCD equations.

The goal of this paper is to derive these analytical solutions and to develop 
their applications in various cases including the important running-coupling 
case. This allows to make a direct and model-independent comparison for 
saturation between QCD predictions and phenomenology.

The plan of the paper is as follows. In Section II, we define consistently 
truncated BK equations for which we derive exact traveling-wave solutions in the 
fixed-coupling case (IIA) and study the validity of the method by comparison 
with numerical simulations of the original BK equation (IIB). In Section III, we 
find solutions to the BK equation with running coupling. In Section IV, we 
apply the results to DIS phenomenology. Conclusions and outlook are given 
in Section V.


\section{The fixed-coupling case}

Let us consider the Fourier transform 
of the dipole scattering amplitude defined as 
\be
{\cal N}(k,Y)=\int_0^\infty 
\f{dr}{r}J_0(kr)N(r,Y)\ .
\label{eq:foutr}\ee
The BK equation for ${\cal N}(k,Y)$ then reads~\cite{kov}
\begin{equation}
\partial_{ Y}{\cal N}=\chi\left(-\partial_L\right){\cal N}- {\cal 
N}^2\ 
\label{eq:bk}
\end{equation}
where $L=\log{\left(k^2/k_0^2\right)}$ and the rapidity $Y$ is measured here in 
units 
of $\bar\alpha\!\equiv\!\ac$ where the strong coupling constant $\alpha_s$ is 
kept fixed.
\begin{equation}
\chi(\gamma)=2\psi(1)-\psi(\gamma)-\psi(1\!-\!\gamma)
\label{eq:kernel}
\end{equation}
is the Balitsky-Fadin-Kuraev-Lipatov (BFKL) kernel \cite{Lipatov:1976zz}. $k_0$ 
is an arbitrary scale.

It has been proved \cite{Munier:2003vc} that the equation \eqref{eq:bk} admits 
asymptotic traveling-wave solutions ${\cal N}(L\!-\!v\ Y)$ where $v$ is the 
speed of the wave. 
Starting with the initial condition ${\cal N}(L,Y_0)\sim\exp{(-\g_i L)},$ 
traveling-wave solutions are formed during the $Y$-evolution 
with the following value for the speed:
\bi
\ii ``pushed front'': $0<\g_i<\g_c; \ \  v= {\chi(\g_i)}/{\g_i},$
\ii ``pulled front'': $\g_i\ge\g_c; \ \   v=v_{c}\equiv {{\chi(\g_c)}/{\g_c}},$
\ei
where $\g_c$ is solution of the equation ${\chi(\g)}/{\g}\!=\!\chi'(\g)$ and is 
called the critical anomalous dimension. Its value is $\g_c\!=\!0.6275$ leading 
to the critical velocity $v_{c}\!=\!4.883.$ Note that \cite{Munier:2003vc} only 
the large$-L$ behavior of the initial condition matters to determine whether one 
lies in a pushed or pulled-front case.

\subsection{Exact traveling-wave solutions of truncated BK equations}

In order to avoid mathematical difficulties related to the infinite-order 
differential equation \eqref{eq:bk}, let us now formally consider a set of 
truncated approximations of the BK equation:
\begin{equation}
 \partial_{ Y}{\cal N}=\chi_{P,\g_0}\left(-\partial_L\right){\cal N}- {\cal 
N}^2\ 
\label{eq:modbk}
\end{equation}
where each truncated kernel $\chi_{P,\g_0}$ is given by the expansion of the 
original kernel $\chi$ around $\g_0$ ($0\!<\!\g_0\!<\!1$) up to order 
$P\!\ge\!2:$
\be
\chi_{P,\g_0}\left(-\partial_L\right)=\sum_{p=0}^P \f{\chi^{(p)}(\g_0)}{p!} 
\left(-\partial_L-\g_0\right)^p=\sum_{p=0}^P (-1)^p A_p
\partial_L^p\ .
\label{eq:modker}\ee
$P$ and $\g_0$ are then two parameters which define a given approximation of the 
original kernel. Each particular kernel $\chi_{P,\g_0}$ is polynomial in 
$\partial_L$ and its 
coefficients $A_p$ can be fully computed from the original kernel $\chi:$ they 
are given by 
\be
A_p=\sum_{i=0}^{P-p}(-1)^i \ \f {\chi^{(i+p)}(\g_0)}{i!\ p!}\ \g_0^i\ .
\label{eq:coeff}
\ee
For the sake of simplicity, we do not write the subscript $P,\g_0$ of the 
coefficients $A_p$. We want to consider kernels $\chi_{P,\g_0}$ that are 
somewhat close to $\chi.$ To be more specific, we shall limit our study to 
kernels that are positive definite and feature $A_0\!>\!0,$ $A_1\!<\!0$ and 
$A_2\!>\!0.$ We also require that, just as the original BK equation 
\eqref{eq:bk}, the equations 
\eqref{eq:modbk} admit asymptotic traveling-wave solutions. In other words, for 
a kernel $\chi_{P,\g_0}$ to be considered, the equation 
${\chi_{P,\g_0}(\g)}/{\g}\!=\!\chi'_{P,\g_0}(\g)\!>\!0$ should have a unique 
solution $\g^{P,\g_0}_c$, which we shall call the critical anomalous dimension. 
We shall also denote the corresponding critical velocity $v^{P,\g_0}_c.$ 

For the kernels $\chi_{P,\g_c},$ for which the expansion has been done 
around $\g_c,$ one has of course $\g^{P,\g_c}_c\!=\!\g_c$ and 
$v^{P,\g_c}_c\!=\!v_c.$ Then to approximate the solutions of the BK 
equation \eqref{eq:bk} by solutions of \eqref{eq:modbk}, one considers
kernels expanded around $\g_0\!\simeq\!\g_c$ to insure that the speed of the 
resulting traveling wave is close to $v_c.$ We also expect that a relevant 
truncation is obtained with a small value of $P.$

Inserting \eqref{eq:modker} into \eqref{eq:modbk}, the truncated BK equation 
reads:
\be
A_0\ {\cal N}-{\cal N}^2-\partial_{ Y}{\cal N}-A_1\ \partial_{ L}{\cal 
N}+\sum_{p=2}^P (-1)^p A_p\ \partial_{ L}^p{\cal N} =0\ .
\label{eq:equiv}\ee
The advantage of this equation w.r.t. the original BK equation is that it can be 
exactly solved \cite{logan,robi} in terms of traveling waves. Indeed, consider 
the 
anzatz
\be
{\cal N}(L,Y)=A_0\ U(s)\ ,
\label{eq:ans}\ee
with the scaling variable $s$ given by
\be
s\equiv\f{\lambda}{c}\ L-\left(A_0+\frac{\lambda}{c}A_1\right)Y
\label{eq:scalvar}\ee
where $\lambda\!=\!\sqrt{A_0/A_2}$ and $c$ is a parameter related to the speed 
of the wave. The equation for the traveling wave $U$ then
becomes the following ordinary differential equation
\be 
U(1-U)+U'+ \f 1{c^2}U''+\sum_{p=3}^P\frac{(-1)^p}{c^p}\ 
\f{\lambda^p\ A_p}{A_0}\ U^{(p)}=0\ .
\label{eq:PU}
\ee
Let us define
\be
v(c)\!=\!A_1\!+\!c\sqrt{A_0A_2}
\label{eq:speed}\ee
such that $s\!=\!\lambda(L\!-\!v(c)Y)/c$ and that $v(c)$ is 
the speed of the traveling wave. For a given kernel $\chi_{P,\g_0}$, the values 
of $c\!>\!-A_1/\sqrt{A_0A_2}$ for which equation \eqref{eq:PU} has a solution 
define the possible values of the speed $v(c)$ of the traveling-wave 
solution \eqref{eq:ans}. 

It is crucial to notice that, in this approach, $c$ is a free and thus 
adjustable parameter. It means that in general, there is a continuum of 
solutions with different speed to the equation \eqref{eq:modbk} or equivalently 
\eqref{eq:equiv}. For instance, if one wants to describe asymptotic solutions,
we are led to choose the value of $c$ such that 
\be
\left\{\begin{array}{ll}v(c)=v_c^{P,\g_0}\equiv\chi'_{P,\g_0}(\g_c^{P,\g_0}) 
&\mbox{if } \g_i\ge \g_c\\
v(c)=\chi_{P,\g_0}(\g_i)/\g_i &\mbox{if } 0<\g_i<\g_c
\end{array}\right.\ . \label{eq:fixc}\ee
Indeed, once the initial condition is fixed, the asymptotic speed is 
$v_c^{P,\g_0}$ or $\chi_{P,\g_0}(\g_i)/\g_i$ depending on the 
initial condition, as confirmed by numerical simulations. We shall use the 
freedom on $c$ to also examine non-asymptotic properties.

Equation \eqref{eq:PU} provides a 
$1/c^p$ expansion in with the $1/c$ term missing. The key point of the method is 
that $1/c$ is indeed a small parameter \cite{logan,robi} allowing to find an 
iterative solution 
\be
h(s) = h_0(s) + \f 1{c^2}h_2(s)+\sum_{p\ge3} \f 
1{c^p}h_p(s) \equiv\f 12 -U(s)\ .\label{eq:series}\ee
Inserting it into \eqref{eq:PU} translates into 
a  hierarchy of equations
\ba
h_0'+h_0^2- 1/4 &=&0\n
h_2'+2h_0h_2+h_0''&=&0\n
h_3'+2h_0h_3-\lambda^3 A_3 h_0'''/A_0&=&0\n
h_4'+2h_0h_4+h_2^2+h_2''+\lambda^4 A_4  h_0''''/A_0&=&0\n
&\cdots &
\label{Phierarchy}
\ea 
where we have written the equations up to ${\cal O}(1/c^5).$ 
An iterative solution can easily be found with initial conditions being 
appropriately chosen, i.e. $h_0(\pm \infty)\!= \!\pm \f 12$ and 
$h_{i\ne 0}(\pm \infty)\! =\! h_{i}(0)\!=\!0.$ Note that if 
$h(s)$ is 
solution, also $h(s\!+\!s_0)$ is solution; this freedom corresponds to the 
possibility of arbitrarily redefining $k_0$. The system is iteratively but fully 
solvable. One first solves the only 
non-linear equation of the hierarchy, obtaining $h_0\!=\!{\scriptstyle \f 12} 
\tanh(\f s{\scriptstyle 2})$. Using the property 
\be
\f {d}{ds}h_n(s)+2h_0 h_n(s) = \f 1{\cosh^2(s/2)}\ \f {d}{ds}\left[\cosh^2(s/2)\ 
h_n(s)\right]\ ,
\label{iteration}
\ee
all the other linear equations reduce to simple integrations of functions 
defined recursively from the  hierarchy. The solution of the first three terms 
of the expansion is 
\begin{equation}
U(s) = \f 1{1\!+\!e^{s}}\!-\!\f 1{c^2}\ \f 
{e^{s}}{\left(1\!+\!e^{s}\right)^2}\  
\log \left[\f {\left(1\!+\!e^{s}\right)^2}{4e^{s}}\right]\! 
-\!\f{\lambda^3}{c^3}\ \f{A_3}{A_0}\ \f {e^{s}}{\left(1\!+\!e^{s}\right)^2}\ 
 \left[3 \f {\left(1\!-\!e^{s}\right)}{\left(1\!+\!e^{s}\right)}\!+\!s\right]+ 
{\cal O}\left(\f 1{c^4}\right)\ .
\label{eq:Psolution}
\end{equation}
The first two terms of the expansion \eqref{eq:Psolution} (order $1/c^0$ and 
$1/c^2$) are {\it universal}: they do not depend neither on $P$ nor on $\g_0.$ 
Indeed 
equation \eqref{eq:modbk} with any kernel $\chi_{P,\g_0}$ admits solutions whose 
first two terms (in the $1/c$ expansion) are those of \eqref{eq:Psolution}. The 
solutions for different kernels only differ through the definition 
\eqref{eq:scalvar} of the scaling variable $s$ which depends on $P$ and $\g_0.$
In this sense, we obtain a universal parametric solution. 

Let us concentrate on the case $P\!=\!2.$ The equation \eqref{eq:modbk} reduces 
to the known FKPP equation and one obtains the corresponding traveling-wave 
equation:
\be
U(1-U)+U'+ \f 1{c^2}U''=0\ .
\label{eq:pe2}\ee
One has $\chi_{2,\g_0}(\g_i)/\g_i\!=\!A_0/\g_i\!+\!A_1\!+\!A_2\g_i$ and
$v_c^{2,\g_0}\!=\!A_1\!+\!2\sqrt{A_0A_2}$ and
therefore, using \eqref{eq:fixc} one 
should take $c\!=\!\lambda/\g_i\!+\!\g_i/\lambda$ in the ``pushed front'' case 
and $c\!=\!2$ in the ``pulled front'' case to reproduce the correct critical 
speed.

In order to check the efficiency of the iterative solution,
we use the existence of an analytic solution~\cite{logan} to equation 
\eqref{eq:pe2}, for the particular value of $c\!=\!5/\sqrt{6}\!\simeq\!2.04:$
\be
U(s)=\left[1+(\sqrt{2}-1)\ \exp\left(\f{cs}{{\scriptstyle 
\sqrt{6}}}\right)\right]^{-2}
\label{eq:exsol}\ .
\ee
This exact solution allows one to see that limiting the expansion after the 
second order is a very good approximation: in Fig.1 we have plotted the exact 
solution \eqref{eq:exsol} along with the solution given by the expansion 
\eqref{eq:Psolution} limited after first and second order with 
$c\!=\!5/\sqrt{6}.$ It is very hard to distinguish the exact solution from the 
solution at second order.

\begin{figure}
\epsfig{file=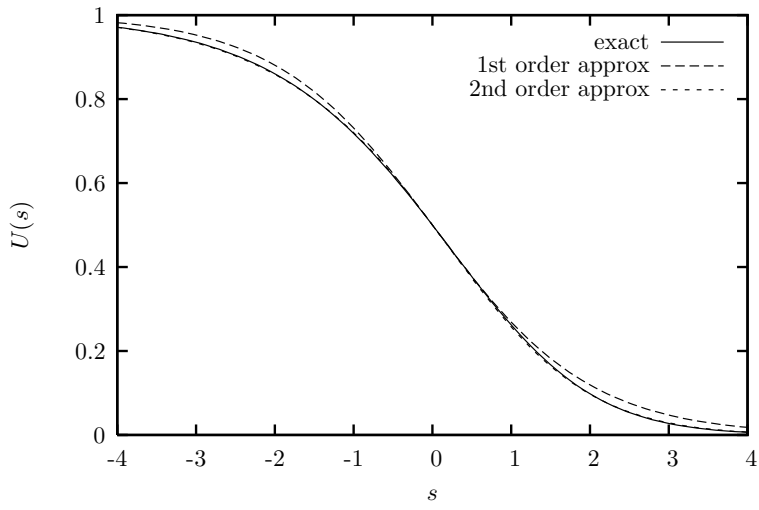,width=10cm}
\caption{The function $U(s)$ solution of the equation $U(1\!-\!U)\!+\!U'\!+ 
\!U''/c^2\!=\!0$ for $c\!=\!5/\sqrt{6}$ with $U(-\infty)\!=\!1$ and 
$U(0)\!=\!1/2.$ Full line: exact solution \eqref{eq:exsol}. Dashed line:   
solution \eqref{eq:Psolution} containing only the first order. Dotted line 
(hardly distinguishable from the full line): solution \eqref{eq:Psolution} 
containing only the first and second order.}
\end{figure}

\subsection{Comparison with numerical solutions of the BK equation}

In terms of physical variables, the solution \eqref{eq:Psolution} of equation 
\eqref{eq:modbk} reads:
\be
{\cal N}(k,Y)=
\f{A_0}{1\!+\!\left[\f {k^2}{Q^2_s(Y)}\right]^{\lambda/c}}\!-\!\f {A_0}{c^2}\ 
\f {\left[\f{k^2}{Q^2_s(Y)}\right]^{\lambda/c}}
{\left(1\!+\!\left[\f{k^2}{Q^2_s(Y)}\right]
^{\lambda/c}\right)^2}\  
\log \f {\left(1\!+\!\left[\f {k^2}{Q^2_s(Y)}\right]^{\lambda/c}\right)^2}
{4\left[\f{k^2}{Q^2_s(Y)}\right]^{\lambda/c}} + {\cal O}\left(\f 1{c^3}\right)
\ ,\label{eq:result}
\ee
where $Q^2_s(Y)\!=\!k_0^2\exp\left[\bar\alpha v(c)Y\right]$ plays the role of 
the saturation scale and where we have restored the $\bar\alpha$ dependence of 
$Y.$ In the following we fix $\bar\alpha\!=\!0.2.$ The $A_p$ coefficients are 
given by \eqref{eq:coeff}, 
$\lambda\!=\!\sqrt{A_0/A_2}$ and $c$ is chosen so that the traveling speed 
$v(c)\!=\!A_1\!+\!c\sqrt{A_0A_2}$ takes the value of the critical speed 
$v_c^{P,\g_0}$ in the ``pulled front'' case and
$\chi_{P,\g_0}(\g_i)/\g_i$ in the ``pushed front'' case, see \eqref{eq:fixc}. In 
the following, we shall only use the first two terms given in \eqref{eq:result} 
which provide a good approximation of the full traveling-wave solution. 

Let us compare numerical solutions of the BK equation with the solution 
\eqref{eq:result} for $P\!=\!2.$ For the numerical simulations, the initial 
condition we chose\footnote{This form ensures correct infrared and ultraviolet 
behaviors with an appropriate matching. The details of this matching are not 
relevant for the asymptotic properties.} is ${\cal N}(L,Y_0)\!=\!e^{-\g_i 
L}/(2\g_i)$ if 
$L\!\ge\!0$ and ${\cal N}(L,Y_0)\!=\!(1\!-\!\g_i L)/(2\g_i)$ if $L\!<\!0.$ As 
already mentioned, only the large$-L$ behavior of the initial condition is 
important to determine whether one lies in the pushed or pulled-front case.
For the analytical solution \eqref{eq:result}, the value of $c$ that gives the 
critical speed is $c\!=\!\lambda/\g_i\!+\!\g_i/\lambda$ in the pushed-front case 
and $c\!=\!2$ in the pulled-front case for which we recall the value 
$\g_c\!=\!0.6275.$ 

The comparison is shown in Fig.2: the ``pulled front'' case is represented on 
the left plot for $\g_i\!=\!1$ and with the kernel expanded around 
$\g_0\!=\!\g_c,$ the critical value. The ``pushed front'' case is represented on 
the right plot for $\g_0\!=\!\g_i\!=\!0.5.$ One sees that as soon as the 
asymptotic traveling-wave regime is reached, there is a good description of the 
``interior'' of the wave, as defined in Ref.\cite{ebert}. In both cases, one 
sees that equation \eqref{eq:result} does not describe neither the saturated 
region (small $k$) nor the ``leading-edge'' (large $k$). Indeed, the exact 
scaling properties of traveling waves are mathematically expected only in the 
interior region \cite{ebert}.

\begin{figure}[htb]
\begin{minipage}[t]{88mm}
\includegraphics[width=8cm]{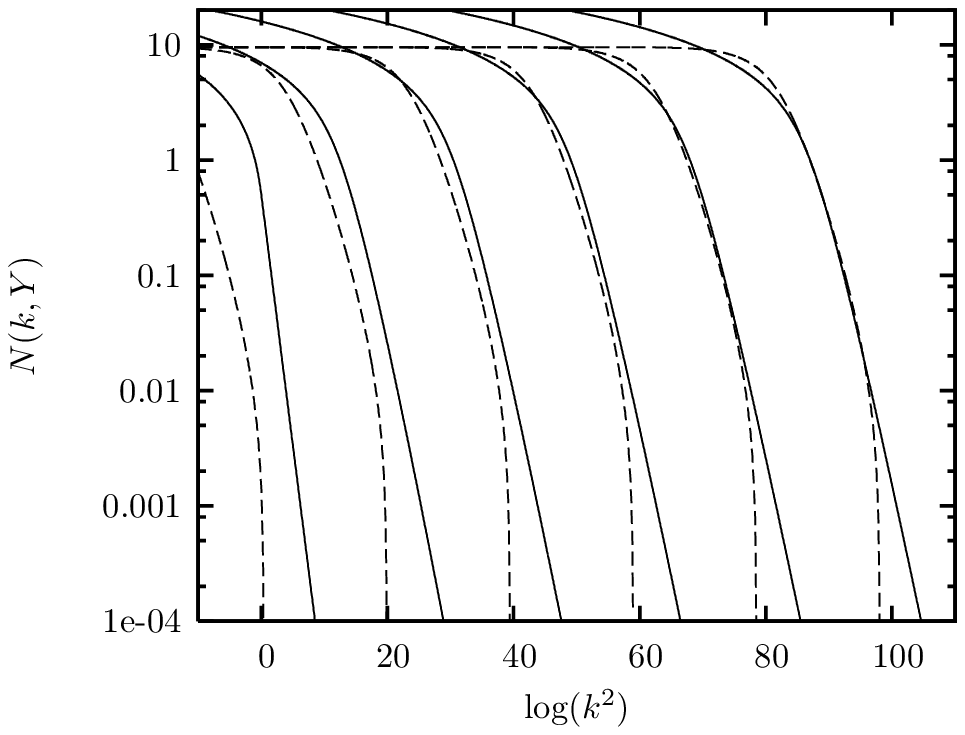}
\end{minipage}
\hspace{\fill}
\begin{minipage}[t]{88mm}
\includegraphics[width=8.2cm]{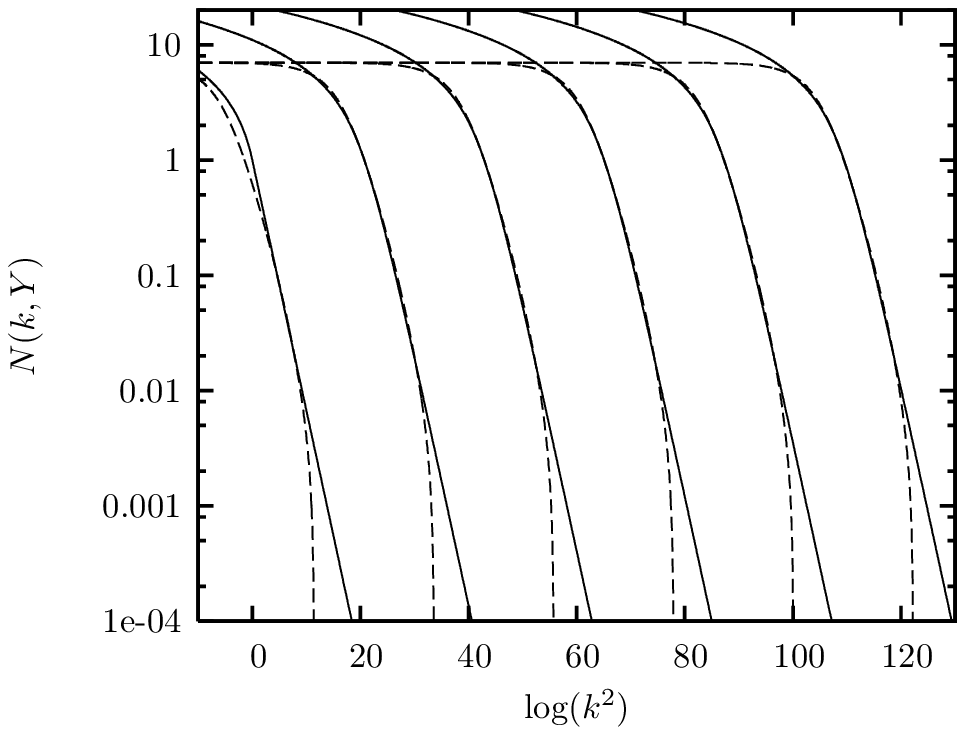}
\end{minipage}
\caption{The traveling wave ${\cal N}(L,Y)$ (fixed coupling) at large $Y$ as a 
function of $L.$ 
The different values of $Y$ are $0,20,40,...,100.$ The full lines are numerical 
solutions of the BK equation and the dashed line are the scaling solutions 
\eqref{eq:result} for $P\!=\!2.$ Left plot, ``pulled front'' case 
($\g_0\!=\!\g_c,$ $\g_i\!=\!1$): the asymptotic traveling-wave regime is reached 
after some evolution with the critical speed $v_c\!=\!4.883.$ Right plot, 
``pushed front'' case ($\g_0\!=\!\g_i\!=\!0.5\!<\!\g_c$): the asymptotic regime 
is driven by the initial condition and the critical speed is 
$\chi_{2,1/2}({\scriptstyle 1/2})/({\scriptstyle 1/2})\!=\!5.55.$}
\label{fig2}
\end{figure}

For our purpose, it is useful to investigate the non-asymptotic regime since it 
is of interest
for phenomenological applications. In Fig.3, we compare the numerical solutions 
of the BK equation with the solution \eqref{eq:result}, in the pulled front 
case, where we adjusted the parameter $c$ to describe the speed of the wave $v$ 
at moderate values of the rapidity Y up to 20. Again the agreement in the 
interior region is satisfactory and it shows that the solution we obtained can 
be used also in the physical range of non-asymptotic rapidities.

\begin{figure}
\epsfig{file=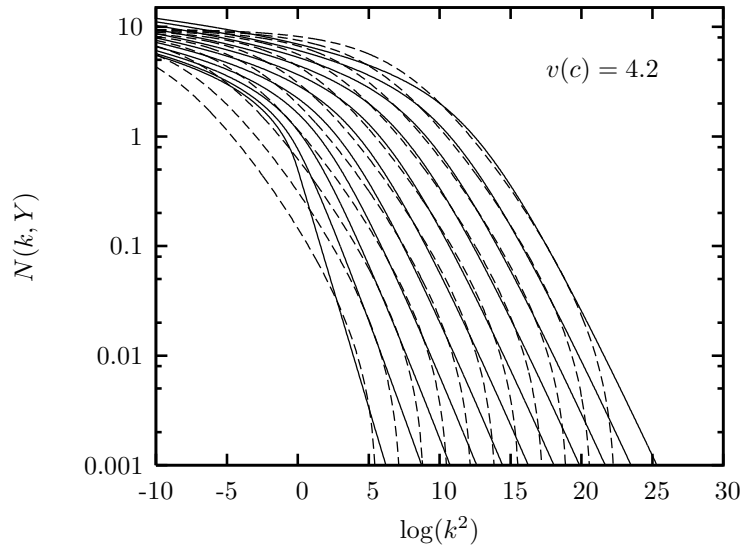,width=10cm}
\caption{The traveling wave ${\cal N}(L,Y)$ (fixed coupling) at moderate $Y$ as 
a function of 
$L.$ The different values of $Y$ are $0,2,4,...,20.$ The full lines are 
numerical 
solutions of the BK equation and the dashed line are the scaling solutions 
\eqref{eq:result} for $P\!=\!2,$ $\g_0\!=\!\g_c,$ and $\g_i\!=\!1.$ The speed of 
the wave has been adjusted to the value $v(c)\!=\!4.2.$}
\end{figure}

Let us finally point out the role of the kernel truncation. First, the 
truncation up to a finite $P$ is mathematically necessary to apply our method. 
Indeed, when considering the full analytic expansion to define the coefficients 
$A_i$ (i.e. $P\to \infty$), they are infinite due to the $\g=0$ singularity of 
the BFKL kernel. This explains why the solutions \eqref{eq:result} describe the 
interior region of the BK front but not the whole $k$ range. For instance, 
one does not describe the saturated region \cite{kozlev}: formula 
\eqref{eq:result} describes 
fronts which saturate at $A_0$ while the BK front behaves as $\log{1/k}$ when 
$k$ is very small. A second remark is that we observed, e.g. by comparison of 
the results of the $P\!=\!2$ and  $P\!=\!4$ truncations, that a good description 
of the interior region of the wave is already obtained for $P\!=\!2$, i.e. the 
diffusive approximation of the BFKL kernel. In this paper, we shall thus stick 
to $P\!=\!2.$


\section{The running-coupling case}

Let us now perform the analysis in the running-coupling case with 
\be
\frac{N_c}{\pi}\ \alpha_s(k^2)=\f1{bL}\ ,\hspace{1cm} b=\f{11N_c-2N_f}{12N_c}\ 
,\hspace{1cm}
L=\log\left(\f{k^2}{\Lambda_{QCD}^2}\right)\ ,
\label{eq:runc}\ee
and $\Lambda_{QCD}\!=\!200\ \mbox{MeV}.$ We thus write the following form of the 
BK equation with running coupling constant and leading-order BFKL kernel 
\eqref{eq:kernel}:
\begin{equation}
bL\ \partial_{ Y}{\cal N}=\chi\left(-\partial_L\right){\cal N}- {\cal 
N}^2\ .
\label{eq:bkrun}
\end{equation}
It has been proved \cite{Munier:2003vc} that this equation admits asymptotic 
traveling-wave solutions of the form ${\cal N}(L\!-\!\sqrt{(2v/b)Y})$ where as 
before $v\!=\!v_c$ in the ``pulled front'' case and $v\!=\!\chi(\g_i)/\g_i$ in 
the ``pushed front'' case.

Let us consider the same truncated kernels than in the previous section. 
That transforms the equation into
\be
A_0\ {\cal N}-{\cal N}^2-bL\ \partial_{ Y}{\cal N}-A_1\ \partial_{ L}{\cal 
N}+\sum_{p=2}^P (-1)^p A_p\ \partial_{ L}^p{\cal N} =0
\label{eq:eqrun}\ee
and we consider again the following form
\be
{\cal N}(L,Y)=A_0\ U(\ts)\ .
\label{eq:ansrun}\ee
Our strategy is to reduce the problem to the one we solved in the previous 
section through an appropriate change of variables. We require that the 
universal terms in the corresponding equation for $U$ are the same as before: 
$U(1\!-\!U)\!+\!U'.$ Postulating $\ts\!=\!L\ \phi(Y/L^2)$ determines a 
solution for $\phi$ and leads to the scaling variable
\be
\ts\equiv
L\ \left(-\f{A_0}{A_1}-\f{1}{\tc}\sqrt{b-2A_1\f{Y}{L^2}}\right)
\label{eq:svrun}\ee
where $\tc$ is a free parameter.
Putting this into \eqref{eq:eqrun} gives
\be 
U(1-U)+U'+\sum_{p=2}^P \left(\f{A_0}{A_1}\right)^p 
\f{A_p}{A_0}\ U^{(p)}+{\cal O}\left(\f 1{\tc}\right)=0\ .
\label{eq:PUrun}
\ee
The leading-order terms in $1/\tc$ are of the same generic form as equation 
\eqref{eq:PU} and thus one can apply the previous method. In this 
running-coupling case, the ${\cal O}(1/\tc)$ terms that we shall neglect contain 
scaling-violation corrections decreasing as $1/L.$ Using the method described in 
the previous Section, one obtains the following solution
\begin{equation}
U(\ts) = \f 1{1\!+\!e^{\ts}}\!-\!\left(\f{A_0}{A_1}\right)^2\f{A_2}{A_0}\
\f{e^{\ts}}{\left(1\!+\!e^{\ts}\right)^2}\  
\log \left[\f {\left(1\!+\!e^{\ts}\right)^2}{4e^{\ts}}\right]\!
+{\cal O}\left(\f{A_0^3}{A_1^3}\right)
\label{eq:solrun}
\end{equation}
where the expansion parameter is $A_0/A_1.$ As before, we shall only 
consider two first terms of the expansion. In order to describe solution to the 
BK equation \eqref{eq:bkrun}, one has now to select the speed of the wave by 
fixing the parameter $\tc.$ Writing at large $Y$
\be
\ts\simeq -\f{A_0}{A_1}
\left(L+\f{A_1}{A_0\tc}\sqrt{-2A_1Y}\right)
\label{eq:expspeed}\ee
and matching with the asymptotic form ${\cal N}(L\!-\!\sqrt{(2v/b)Y}),$
one gets
\be
v(\tc)=-\f{bA_1^3}{A_0^2\ \tc^2}\ .
\ee
Adjusting $\tc$ to obtain a given speed, we check that $1/\tc$ is indeed small. 
Defining 
$\tilde{Q}_s^2(Y)\!=\!\Lambda_{QCD}^2\exp(\sqrt{(2v(\tc)/b)Y})$ that plays the 
role of the saturation scale, the scaling variable reads in terms of physical 
variables
\be
\ts=-\f{A_0}{A_1}\log\left(\f{k^2}{\Lambda_{QCD}^2}\right)
-\f{1}{\tc}\sqrt{b\log^2\left(\f{k^2}{\Lambda_{QCD}^2}\right)
+\f{A_0^2\ \tc^2}{A_1^2}\log^2\left(\f{Q_s^2(Y)}{\Lambda_{QCD}^2}\right)}\ .
\label{eq:varrun}\ee
Formula \eqref{eq:varrun} along with
\begin{equation}
{\cal N}(k,Y)= \f {A_0}{1\!+\!e^{\ts}}\!-\!\f{A_0^2A_2}{A_1^2}\
\f{e^{\ts}}{\left(1\!+\!e^{\ts}\right)^2}\  
\log \left[\f {\left(1\!+\!e^{\ts}\right)^2}{4e^{\ts}}\right]
\label{eq:parrun}
\end{equation}
is the traveling-wave solution of the truncated BK equation with running 
coupling \eqref{eq:bkrun} at ${\cal O}(1/\tc).$

Let us compare numerical solutions of the full BK equation
\eqref{eq:bkrun} with that parametrization. For the numerical simulations, the 
initial condition is the same as in the fixed-coupling case. The problem of the 
Landau pole of the running coupling is dealt with by introducing a regulator: 
$\log(k^2/\Lambda_{QCD}^2)\!\rightarrow\!\log(\kappa\!+\!k^2/\Lambda_{QCD}^2)$ 
with $\kappa\!=\!3.$ As well as the details of the initial condition, this does 
not modify the traveling-wave pattern.

The comparison is represented in Fig.4 where the 
``pulled front'' case is considered. The values of the 
coefficients $A_0\!=\!9.55,$ $A_1\!=\!-25.6$ and $A_2\!=\!24.3$ have been fixed 
by considering $P\!=\!2$ and $\g_0\!=\!\g_c.$
We use the free parameter $\tc$ to adjust the speed of the wave.
On the left plot, high values of the rapidity $Y\!=\!0,10,20,...,100$ are 
considered while on the right plot, moderate values $Y\!=\!0,2,4,...,20$ are 
represented. In both cases, in order to describe the numerical solutions, one 
has to adjust the actual speed of the wave $v(\tc)$ to the value of $3.1$ for 
the left plot and $2.3$ for the right plot which are both significantly smaller 
than the critical value $v_c\!\simeq\!4.88.$ It confirms that the establishment 
of the asymptotic regime is much slower in the running-coupling case than in the 
fixed coupling case \cite{Munier:2003vc}. Indeed, the theoretical sub-asymptotic 
corrections to the critical speed have been computed \cite{Munier:2003vc} and 
predict such a behavior. There is a good description of the interior of the wave 
for the different ranges of rapidity, provided we adjust the speed of the wave.  
As expected, the parametrization does not describe the saturated and the 
leading-edge regions. Note that the scaling region at low values of $Y$ is less 
extended in the running coupling case (Fig.4b) than for fixed coupling (Fig.3), 
this could be due to the scaling-violation terms that we had to neglect in 
\eqref{eq:PUrun}. However it is clear that an approximative traveling-wave 
pattern already emerges at moderate rapidities.


\begin{figure}[htb]
\begin{minipage}[t]{88mm}
\includegraphics[width=8.8cm]{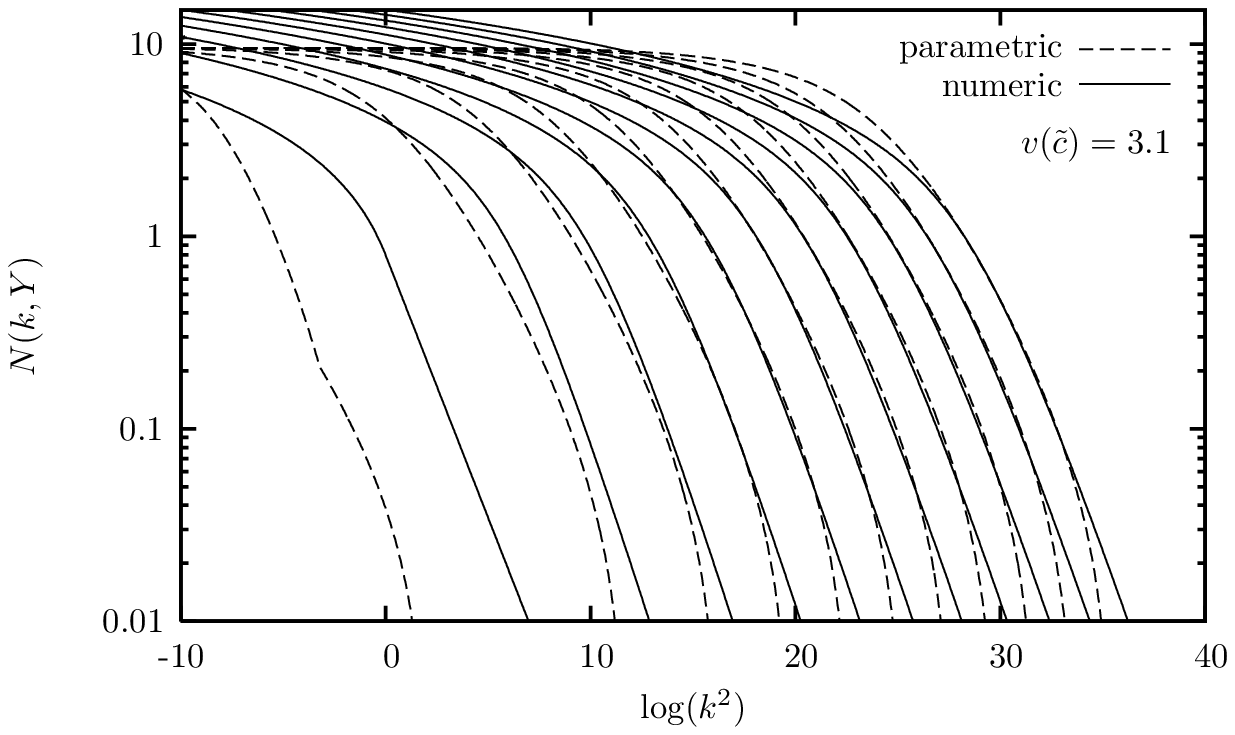}
\end{minipage}
\hspace{\fill}
\begin{minipage}[t]{88mm}
\includegraphics[width=8.8cm]{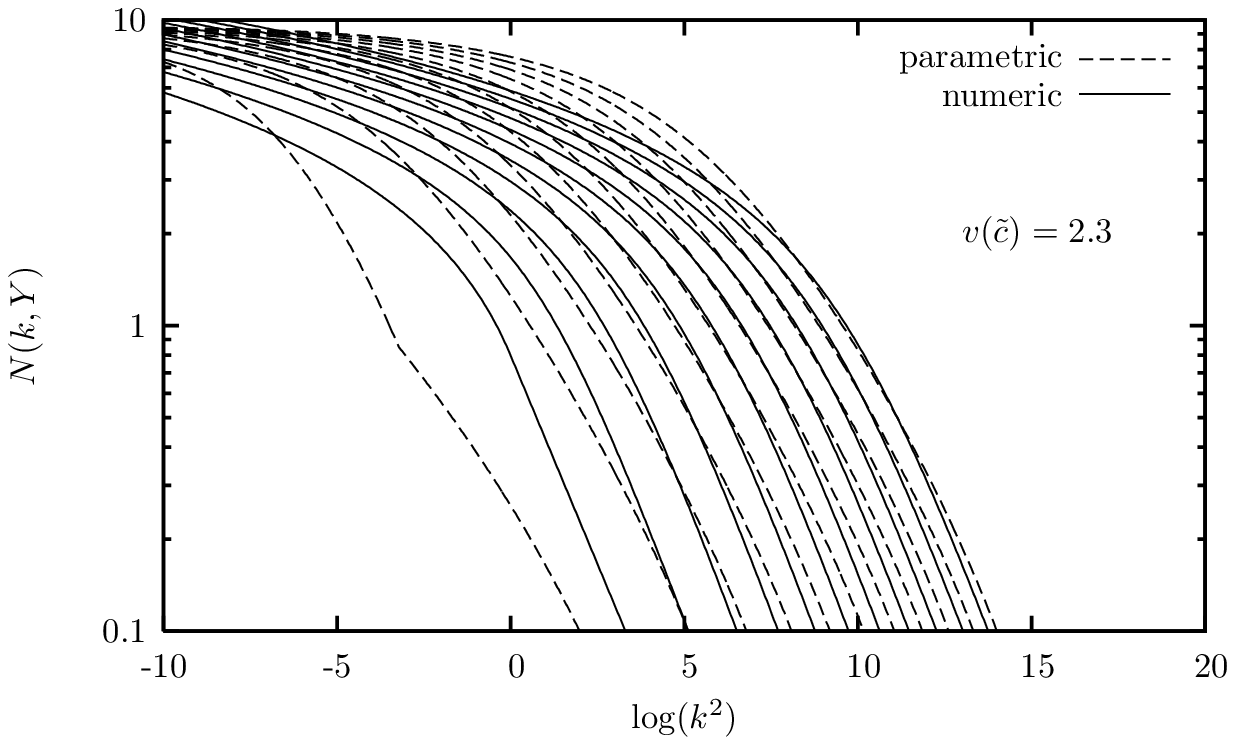}
\end{minipage}
\caption{The traveling wave $N(L,Y)$ with running coupling as a function of $L.$ 
The full lines are numerical 
solutions of the BK equation with running-coupling and the dashed lines are the 
parametrization \eqref{eq:varrun} and \eqref{eq:parrun}. Left plot: high-energy 
regime $Y\!=\!0,10,20,...,100$ and $v(\tc)\!=\!3.1.$ Right plot: moderate energy 
regime $Y\!=\!0,2,4,...,20$ and $v(\tc)\!=\!2.3.$ The initial condition is 
chosen 
in the pulled front regime.}
\label{fig3}
\end{figure}

\section{Application to deep inelastic scattering}

We shall now draw a link between traveling waves and phenomenology. The 
observation of geometric scaling asks the question whether it can be viewed as a 
consequence of QCD traveling waves. The known problem is that the range of 
rapidity for which asymptotic predictions of QCD non-linear evolution equations 
are available remains far from the moderate rapidity range accessible to 
experiments. Hence the phenomenological interest of our method is to open a way 
of confronting theory and experiment in some physical kinematic range.

The formula relating the dipole scattering amplitude in coordinate space 
$N(\mathbf{r},\mathbf{b},Y)$ to the unintegrated gluon distribution of the 
target $f(Y,k^2)$ is (see e.g. 
\cite{Bialas:2000xs}):
\be
\int d^2b\ N(\mathbf{r},\mathbf{b},Y)=\f{4\pi^2\alpha_s}{N_c}
\int \f{dk}{k}\ f(Y,k^2)(1-J_0(k|\mathbf{r}|))\ .
\label{eq:defun}\ee
In the $\mathbf{b}-$independent approximation, one writes $\int d^2b\ 
N(\mathbf{r},\mathbf{b},Y)=\pi R_p^2 N(r,Y)$ where $R_p$ is the radius 
of the proton target. Through formulae \eqref{eq:foutr} and \eqref{eq:defun} a 
link is established between ${\cal N}(k,Y)$ and $f(Y,k^2).$ One gets
\be
{\cal N}(k,Y)=\f{4\pi\alpha_s}{N_cR_p^2}\int_k^\infty 
\f{dp}{p}\ f(Y,p^2)\log{\left(\f{p}{k}\right)}\ .
\ee
Finally, we use the relation
\be
f(Y,k^2)=\f{\partial}{\partial k^2}\ xg(x,k^2)\ ,
\label{eq:rel}\ee
where $Y\!=\!\log(1/x)$ between the unintegrated gluon distribution $f(Y,k^2)$ 
and the gluon distribution $xg(x,k^2)$ to obtain
\be
{\cal N}(k,Y)=\f{\pi\alpha_s}{N_cR_p^2}\int_{k^2}^\infty 
\f{dt}{t^2}\ xg(x,t)\log{\left(\f{t}{ek^2}\right)}\ .
\label{eq:pdf}\ee
Note that formula \eqref{eq:rel} is a well-known approximation which may only be 
valid at small values of $x.$ It is useful here as it makes the analysis 
simpler. For a more detailed study which is beyond the scope of this work, one
could consider more advanced prescriptions \cite{martin}.

In order to compare \eqref{eq:pdf} to our predictions, we use the 
running-coupling case which is known to lead to traveling waves with a speed 
compatible with phenomenology \cite{dyonisis}. 
\begin{figure}
\epsfig{file=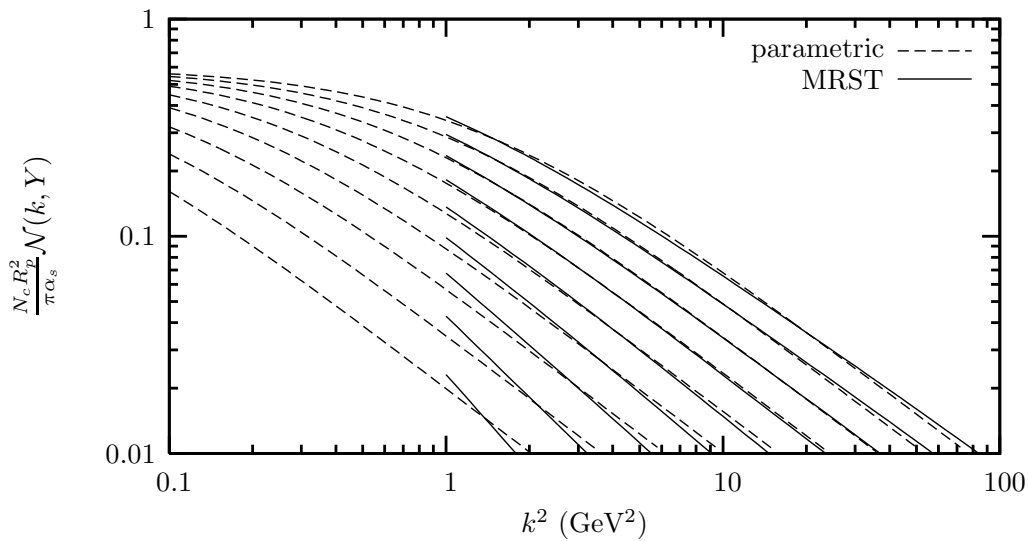,width=14cm}
\caption{The dipole amplitude $N(k,Y)$ as a function of $L$ for different values 
of $Y\!=\!4,5,...,12.$ The full lines are obtained from the MRST gluon 
distribution via \eqref{eq:pdf} and the dashed lines represent the 
parametrization \eqref{eq:varrun} and \eqref{eq:parrun}. Only one term is kept 
in \eqref{eq:parrun}.}
\end{figure}

We performed a fit of the parametrization \eqref{eq:varrun} and 
\eqref{eq:parrun} on the amplitude \eqref{eq:pdf} obtained from the 
Martin-Roberts-Stirling-Thorne (MRST) gluon distribution \cite{mrst}. The 
domain we considered is $x\!<\!10^{-3}\!$ and ${\scriptstyle 
\f{N_cR_p^2}{\pi\alpha_s}}{\cal N}(k,Y)\!>\!0.01$ which is the region where 
traveling-wave patterns can be tested. 
Note that we considered only the first term of \eqref{eq:parrun} which depends 
on the parameters $A_0,$ $A_1,$ and $v$ the speed of the wave. We discarded the 
second term of \eqref{eq:parrun} since we found that the parameter $A_2$ stays 
undetermined by the fit in the kinematical region we considered. In Fig.5, we 
show the result of the fit, the obtained parameters being $A_0\!=\!17.1,$ 
$A_1\!=\!-15.8,$ and $v\!=\!1.76.$ 

These results call for comments.
\vspace{-0.1cm}
\begin{rmenum}
\item The dipole amplitude \eqref{eq:pdf} 
obtained from the MRST distribution (full lines on Fig.5) displays a structure 
compatible with an evolution towards traveling waves, namely a steeper slope at 
small rapidities which evolves toward a less steep and regular pattern at higher 
rapidities.
\item Fig.5 shows a compatibility with the parametric form (\eqref{eq:varrun}, 
\eqref{eq:parrun}) at high values of $Y$ and moderate values of $k^2.$ 
\item On the theoretical ground, it is worth comparing the fitted parameters to 
various predictions. For the truncated BK kernel with $P\!=\!2$ and 
$\g_0\!=\!\g_c,$ one finds $A_0\!=\!10,$ $A_1\!=\!-25,$ and the speed
$2.3\!<\!v\!<\!3$ when one includes pre-asymptotic corrections 
\cite{Munier:2003vc}. 
With $P\!=\!4$ and $\g_0\!=\!1/2,$ one finds $A_0\!=\!11,$ $A_1\!=\!-49.$ 
The obtained parameters are qualitatively closer to the diffusive approximation 
($P\!=\!2$) of the BK equation but they are quantitatively different. This seems 
to suggest the presence of next-leading or higher-order corrections to the QCD 
kernel. This deserves further study.
\item Note that we obtain $|A_0/A_1|\!\sim\!1$ for the observed parameters which 
could invalidate the expansion \eqref{eq:solrun}. We checked both numerically 
and from the analytical form that the second and higher order terms stay 
sufficiently small to be still neglected.
\end{rmenum}

\section{Conclusion}

Let us summarize the main results of our study.\\
- We found iterative traveling-wave solutions to non-linear BK evolution 
equations 
obtained by finite truncation of the BFKL kernel for both fixed and running 
coupling constant.\\
- These solutions exhibit universality properties, the first two dominant terms 
of the iteration have a parametric form independent of the truncation of the 
kernel and of the fixed or running coupling cases.\\
- The scaling variable is found to be a combination of $L\!\equiv\!\log(k^2)$ 
and rapidity $Y$ which is given by formula \eqref{eq:scalvar} (resp. 
\eqref{eq:svrun}) for fixed (resp. running) coupling.\\
- The obtained traveling-wave solutions match with the asymptotic solutions of 
the BK equation. This was verified by analytical and numerical checks. The 
remarkable new property is that they also match with the non-asymptotic behavior 
of the interior of the wave, provided an adjustment of the speed of the wave 
which is a free parameter in the iterative approach.\\
- As an application of the method and its validity at non-asymptotic energies, 
we 
considered the dipole amplitude in momentum space ${\cal N}(k,Y)$  obtained from 
the MRST parametrization of the gluon distribution. We found evidence for an 
evolution pattern compatible with the formation of traveling waves. The obtained 
dipole amplitude is well described by the universal parametric form 
\eqref{eq:parrun}. The obtained parameters seem to point towards traveling-wave 
solutions for a BFKL kernel modified by higher orders and/or non-perturbative 
corrections.

Our results help establishing a stronger link between the theory of 
saturation and the phenomenological observation of geometric scaling 
\cite{Stasto:2000er}. Moreover, simple analytical parametrizations of the 
solutions of the BK equation in momentum space could help extending the present 
phenomenology \cite{pheno} to a broader range of kinematics or observables. It 
would be interesting to investigate where one could distinguish between a DGLAP 
structure and traveling waves. On a more theoretical level, it 
seems feasible to extend the method to  QCD evolution equations beyond the 
``mean field''  BK equation  (\ref{eq:bk}). In particular the extension to the 
stochastic versions of  QCD evolution equations \cite{stochastic} would help our 
understanding of QCD in the high-energy limit.

\begin{acknowledgments}
G.S. is funded by the National Funds for Scientific Research (Belgium).
\end{acknowledgments}


\end{document}